\documentclass[aps, twocolumn, pra, showpacs,superscriptaddress]{revtex4}
\usepackage{graphicx}
\usepackage{amsmath}
\usepackage{amsfonts}
\usepackage{amssymb}
\usepackage{epsfig}

\begin{document}

\title{An optical example for classical Zeno effect}
\author{Li-Gang Wang}
\affiliation{Department of Physics and ITP, The Chinese University of Hong Kong, Hong
Kong, China}
\affiliation{Department of Physics, Zhejiang University, Hangzhou 310027, China}
\author{Shi-Jian Gu}
\affiliation{Department of Physics and ITP, The Chinese University of Hong Kong, Hong
Kong, China}
\author{Hai-Qing Lin}
\affiliation{Department of Physics and ITP, The Chinese University of Hong Kong, Hong
Kong, China}

\begin{abstract}
In this brief report, we present a proposal to observe the classical zeno
effect via the frequent measurement in optics.
\end{abstract}

\pacs{05.40.-a, 03.65.Xp, 03.67.-a}
\date{\today }
\maketitle



Zeno paradox, such as the problem of the so-called flying arrow, contrary to
the evidence of our senses, seems never happen in real classical world. But
this paradox is believed to be possible realized uniquely in quantum world,
known as quantum Zeno effect proposed by Misra and Sudarshan \cite{BMisra77}
in 1977, because of probability properties of quantum states and the
projective measurement in quantum mechanics (for a review, see Ref. \cite%
{KKoshinoPhysRep}). Recently, one of us (Gu) argued that the classical Zeno
effect is possible recovered with Super Mario's intelligent feedback \cite%
{SJGuMario}. Later, we further showed that the decay of a classical state in
classical noise channels can be significantly suppressed with the aid of the
successive repeaters \cite{SJGuRepeater}, in this sense we claim that the
classical Zeno effect may exist in classical stochastic process. In this
report, we present a proposal to observe the classical zeno effect in
optics. The evolution of the polarized-light intensity in the designed
system are strongly affected by the measured times.

\begin{figure}[tbp]
\centering
\includegraphics[width=8cm]{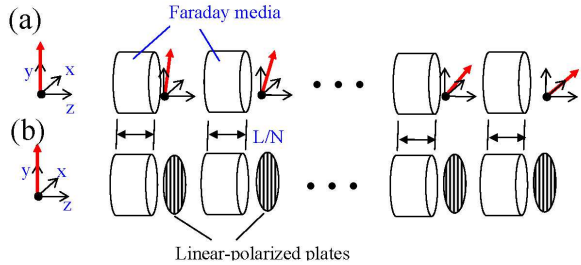}
\caption{(Color online) (a) Setup for successive polarization rotation with
a series of Faraday media and (b) setup for the observation of optical Zeno
effect with vertical-polarization measurements after each Faraday medium.
Each Faraday medium with the length $L/N$ induces a polarization rotation
angle of $\protect\pi /2N$ and $L$ is the total length of all Faraday media.
}
\label{fig:FIG1}
\end{figure}

As shown in Fig. 1(a), when a linear polarized light beam with initial
intensity $I_{0}$ is incident on a series of successive Faraday media, the
polarization direction of the beam gradually changes with the increasing
number of the Faraday media. Assuming that the initial direction of light
polarization is in the $y$ direction and the angle for the polarization
rotation from the input to output changes $\pi /2$, then the intensity of
the linear polarization beam for the $y$ component is given by%
\begin{equation}
I(z)=I_{0}\cos ^{2}\left[ \frac{\pi z}{2L}\right] ,
\end{equation}%
where $L$ is the total length of all Faraday media that induce the $\pi /2$
rotation of light polarization, and $z$ is the internal distance inside the
Faraday media. Since there is no measurement in (a), the intensity of
polarization beam for the $y$ component evolutes smoothly as a function of $%
z $.

\begin{figure}[tbp]
\centering
\includegraphics[width=8cm]{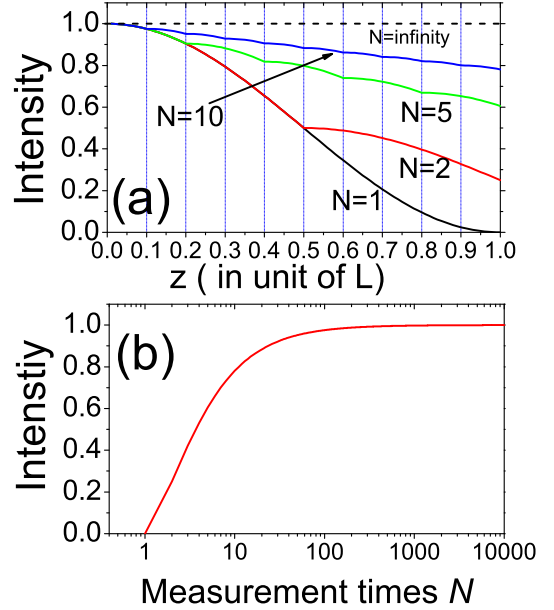}
\caption{(Color online) (a) Intensity evolutions of the polarized beam in
the $y$ component as a function of distance $z$ for different measurement
times. (b) Dependence of the output intensities of the polarized beam on the
measurement times $N$. }
\label{fig:FIG2}
\end{figure}

However, once the measurements are presented in Fig. 1(b), the evolution of
the resulted intensity is dramatically changed. In Fig. 1(b), we do a
vertical-polarization measurement (along the $y$ direction) after each
Faraday medium, then the resulted intensity of the polarization beam for the
y component becomes

\begin{equation}
I(z)=I_{0}\left[ \cos ^{2}(\frac{\pi }{2N})\right] ^{i-1}\cos ^{2}\left\{
\frac{\pi }{2L}\left[ z-\frac{L}{N}\left( i-1\right) \right] \right\} ,
\end{equation}%
where $N$ is the total number of Faraday media with the same polarization
rotation angle $\pi /(2N)$, and $i$ denotes the $i$th Faraday medium at
which the distance z is located. Therefore, the intensity for the
polarization light beam for the y component at the output end finally becomes

\begin{equation}
I_{\text{out}}=I_{0}\left[ \cos ^{2}\left( \frac{\pi }{2N}\right) \right]
^{N}.
\end{equation}%
When the measured times increase to be infinite, i. e., $N\rightarrow \infty
$, the output intensity at the output end will be close to $I_{0}.$ This
indicates that the initial intensity of the linear polarized light beam for
the y component, passing through numerical Faraday media with small
polarization rotation angles, will not decay after the infinite
measurements. Figure 2(a) shows clearly the changes of the intensities of
the linear polarized light in the $y$ component for different measurement
times, and with the increasing of the measurement times the decay of the
intensity becomes slower. In Fig. 2(b), it is found that the output
intensity at the output end becomes larger and larger, and it gradually
tends to be one with the increasing of the measurement times $N$.

In summary, instead of the memory effect in the previous scheme \cite
{SJGuRepeater}, in the present scenario we use the polarization property of
light to recover the Zeno effect. We can see that the measurement of the light
polarization plays the same role of the projective measurement in quantum
mechanics. However, all the quantities involved here are classical. In a word,
Zeno effect does happen in the classical world.

\begin{acknowledgments}
This work is supported by the Earmarked Grant Research from the Research
Grants Council of HKSAR, China (Project No. CUHK 400807 and 403609).
\end{acknowledgments}

\end{document}